# OSD: A Source Level Bug Localization Technique Incorporating Control Flow and State Information in Object Oriented Program

Partha Pratim Ray
IEEE Student Member
Cooch Behar, West Bengal, India
parthapratimray@hotmail.com

*Abstract*—Bug localization in object oriented program has always been an important issue in software engineering. In this paper, I propose a source level bug localization technique for object oriented embedded programs. My proposed technique, presents the idea of debugging an object oriented program in class level, incorporating the object state information into the Class Dependence Graph (ClDG). Given a program (having buggy statement) and an input that fails and others pass, my approach uses concrete as well as symbolic execution to synthesize the passing inputs that marginally from the failing input in their control flow behavior. A comparison of the execution traces of the failing input and the passing input provides necessary clues to the root-cause of the failure. A state trace difference, regarding the respective nodes of the ClDG is obtained, which leads to detect the bug in the program.

## I. INTRODUCTION

From the last few years embedded systems have established itself as unavoidable criteria in human society. Due to its low code size and less complexity embedded systems are being implemented in the most sophisticated and critical applications. With this advent of havoc implication of em bedded systems, the whole science community is now moving towards object oriented methods to fulfill the excessive need of these systems. The power of handling complexity is the added advantage to the object oriented technologies that enable them to compete the other traditional techniques like as procedural approach.

Debugging denotes the process of detecting root causes of unexpected observable behavior in programs (such as a program crash, an unexpected output value being produced or an assertion violation). Debugging program errors is a difficult process, and often takes a significant fraction of the time in the program development stage. Even today, debugging remains much of a manual activity, with the actual debugging time dependent on the size and complexity of the program being debugged, the nature of manifestation of the bug and the level of familiarity and expertise of the programmer. The standard practice of debugging till date in the software community is to manually inspect the execution trace exhibiting the bug inside a debugger and try and locate the error cause(s) from an observed error.

In the past decade, there have been several attempts to automate the debugging activity by fully automated / semi-automated formal analysis of the program and/or the failed execution trace for software programs. These methods, in spite of rich theoretical foundations and promising automated bug finding capabilities, have found a low degree of acceptance and penetration in the research and industrial community till date. The main challenge is to develop a scalable solution that can handle softwares of sizeable complexity and pin-point the root cause(s) of an observed error with a high level of accuracy.

Each software needs to undergo a very crucial stage of its life cycle-- debugging process. Whenever a program behaves unexpectedly thus producing wrong output is liable to be called a buggy program. In effect to remove the bug from the program the debugging methodology should be very stable one. Though different techniques are already available to debug an object oriented program, they all are not very suitable for the targeted problem, such as having a prominent state chart in form of UML. In this type of cases I need to imply a new technique that adds object state information of the class being executed, into the ClDG. This helps in knowing the root cause of the bug, introduced in the program under execution.

The *Class Dependence Graph* (ClDG) represents the control and data dependencies within a class [1]. For a given class, the ClDG consists of a set of *program dependence graphs* (PDGs) [2] with additional edges to represent inter-procedural control and data dependences. A statement in a procedure is represented by a statement *vertex*. Control and data dependences between program statements are represented by *control dependence* and *data dependence* edges, respectively. In this paper I first take a buggy object oriented program and generate a state chart UML diagram and ClDG. After the models are generated I input some test cases into the buggy program, that results a fail and pass traces. Then one of the pass cases is selected the match class dependence flow to the failed one. This results in object state comparison between the pair of pass and fail cases. Hence producing the bug report telling the position of bug inside the buggy program.

This paper is organized as follows: Section II presents related work. Section III presents an overview of my approach. Section IV presents detailed methodology, while Section V ends with conclusion.

## II. RELATED WORK

Many testing technique has been already proposed in literature for testing in traditional programs [4]. Literature [7] tells about the selection of regression tests in object oriented programs. [8] indentifies the test coverage requirements for modified softwares. In [3], a model based regression test selection methodology has been presented. Existing model-based approaches for traditional programs [9] construct graphical models based solely on source code analysis of the programs. [12] proposes real time software debugging technique. [13] tells about exception handling in respect to testing of software. [14] also tells some improvement on exception handling in testing area. [16] tells about the state chart based object oriented integrated testing. In [17] different aspects of testing and debugging has been presented well. Different search algorithms for regression test cases has been proposed in [18].

## III. OVERVIEW OF OSD APPROACH

I have named my technique Object State-based *Debugging* for object oriented program (OSD). My technique is essentially based on first constructing models for buggy program that is state chart and ClDG. Then state transition table is created according to the state chart. Here one thing is to be noted that ClDG is incorporated with the state chart. That means each node of ClDG is accompanied with a state (state of object of same class). Then a test suite $t_1,...,t_i,...t_j,...t_k$, is applied as input into the buggy program. This results in some pass and some fail cases. Suppose the set of pass cases is $t_1,...,t_{j-1},t_{j+1},...,t_k$ and the fail set is $t_j$ (single element). Now at this point OSD, control dependence flow comparison is made between fail set-$t_j$ and all of pass set-$t_1,...,t_{j-1},t_{j+1},...,t_k$. After the comparison, $t_i$ is the sole test input that matches the most to $t_j$ (in respect to control dependence flow). At this stage of OSD State comparison occurs between $t_i$ and $t_j$, resulting bug report that points out the bug in source code level.

The important steps of my approach have been shown in Figure 1; the rounded corner blocks represent initial input test suite, control dependence flow comparison, state comparison, bug report etc. The rectangular blocks represent buggy object oriented program, ClDG, state chart UML, state transition table and other output test case and sets etc. I now briefly discuss the different steps involved in my approach.

1) The *Buggy object oriented program* is the source code written in C++ language. It is basically a buggy code where the logic of an elevator controller (embedded system) is illustrated. My target is to debug this code.
2) The *State chart UML* is the finite state machine model corresponding to the buggy program.
3) The *State transition table* is generated from state chart UML model. It contains various initial and final states with their condition and operation embedded.

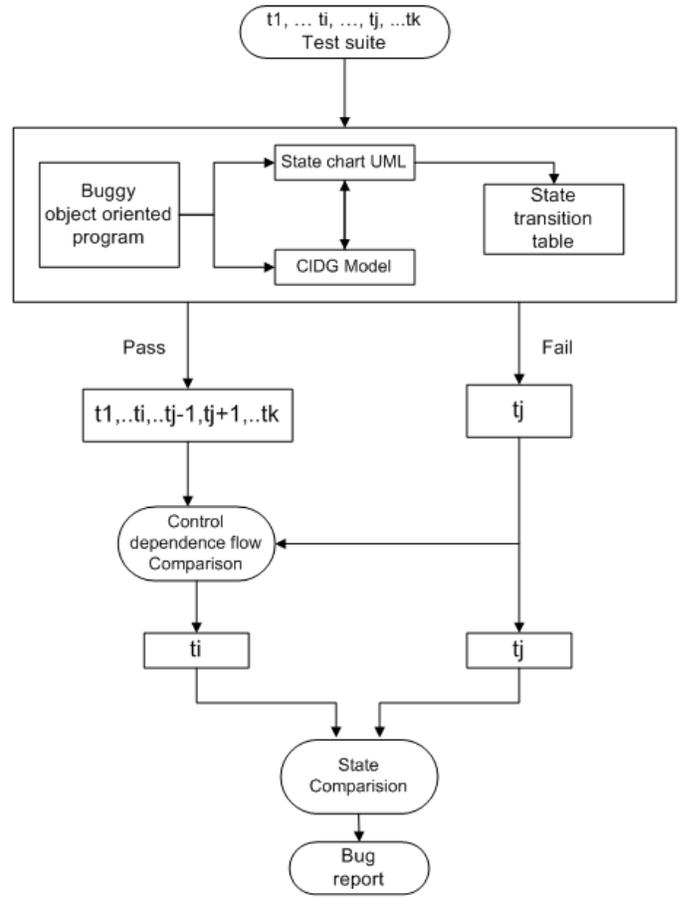

Fig. 1. Overview approach to OSD

4) The *ClDG* model represents the class dependence graph of the buggy program itself.
5) The *Control dependence flow comparison* block compares the control dependence flow between the failed and all other passed input.
6) The *State comparison* block compares the states of $t_i$ and $t_j$, between the nodes of ClDG.
7) The *Bug report* is the final report regarding bug present inside the buggy program, which can point out the location of bug in source code level.
8) The $t_1,...,t_i,...,t_j,...,t_k$ is the input test suite.
9) The $t_1,...,t_i,...t_{j-1},t_{j+1},...,t_k$ is the pass input of the previous told suite.
10) The $t_j$ is the failed input from the input suite.
11) The $t_i$ is the best chosen input from the set of pass inputs belonging to initial test suite.

## IV. DETAILED APPROACH

In this section I will describe the detailed methodology of Object State based Debugging (OSD). First thing that I will present in this portion is the buggy program.

## A. Buggy Object Oriented Program

The Figure 2 is the code snippet of an elevator controller written in C++. The controller has two main parts.

- *Request Resolver* – resolves various floor requests into single requested floor.
- *Control* – moves elevator to its requested floor.

*1) Description:* Elevator [11] moves either up or down to reach the requested floor. Once at the requested floor, open the door for at least 10 seconds, and keep it open until the requested floor changes. It is ensured that the door is never open while moving. Elevator does not change directions unless there are no higher requests when moving up or no lower requests when moving down. In this paper, I have taken the building to be three storied (having number of floors equal to three that is - ground -0, first-1, second-2) for simplicity.

*2) Prefixes introduced in program:* The numbers have been assigned sequentially to each statement in the order they appear in the source code for identifying them in the ClDG. The prefixes S, E, CE and C denote statements, method entry, class entry and call nodes respectively.

*3) Bug intoduced:* In respect to my investigation I have introduced a bug inside the code. Line number 13 and 14 in unitControl() method of Control class. This results in *door open* when any one, requests a floor (which is 1 in this case) from a lower floor (say floor number 0). This repels the door to be open for 10000 milliseconds, thus not invoking the expected task (movement). This restricts the person (standing at ground floor) to go first floor. This code works well otherwise.

*4) Concurrency error ignorance:* This code snippet is purely an example of concurrency. Hence the hazards regarding concurrency such as deadlock, synchronization etc. have been ignored at time.

```
     int up, down, open, floor=0, req;
CE1  class RequestResolver{
E2       int resolver(){
S3           while (1){
                 ...
                 //Get request from user
S4               cin>>req;
             }
     };
CE5  class Control{
E6       void unitControl(){
S7           int time= up = down = 0,open = 1;
S8           while (1) {
S9               while (req == floor); // Idle
S10                  open = 0;
S11              if (req > floor) { up = 1; // Going Up
S12                  while (req != floor) {
S13                      if (req == 1)
S14                          goto P;
S15                      floor++; } }
S16              else {down = 1;  // Going Down
S17                  while (req != floor)
S18                      floor--; }
S19          P: up = down = 0; open = 1;
                 // Wait for 10000 ms // Door Open
S20              for ( time=0; time < 10000; time++);}
             }
     };
E21  void main()
     {
S22      Control *c = new Control();
S23      RequestResolver *r = new RequestResolver();
S24      while(1) {   // Call concurrently:
C25          r -> resolver();
C26          c -> unitControl();}
     }
```

Fig. 2. Buggy Object Oriented Program snippet

## B. State chart

I have obtained a state chart UML diagram from Figure 2. The diagram is shown in the Figure 3. The state chart shows four different states – *Idle, Going Up, Going Down, Door Open* as the probable states. Possible transitions from one state to another is based on input (E.g., req > floor, req < floor etc.). Actions are occurred in each state (E.g., the GoingUp state *u,d,o,t = 1,0,0,0* (up = 1, down, open, and timer_start = 0).

## C. State transition table

Table I refers to the state transition table generated from Figure 3. The transition table contains initial state, condition, operation / action, final state. There is another state `Not Defined`– ND, that can play an important role in providing object state information to those states which can always not satisfy the given state chart criteria. Going up = u, Going Down = d, Door open = o, timer start = t are different notions of actions, that frequently occur in each state of the class (object state information) under execution.

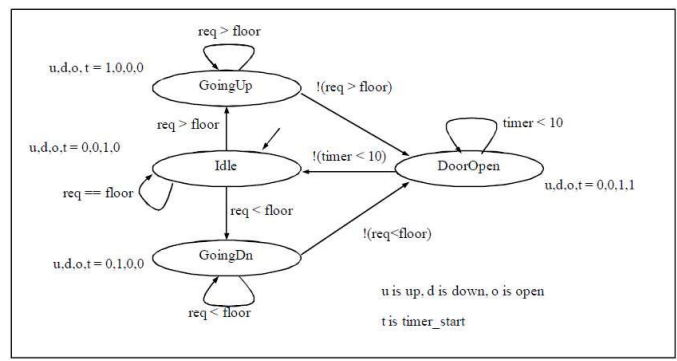

Fig. 3. State chart generated from Figure 2

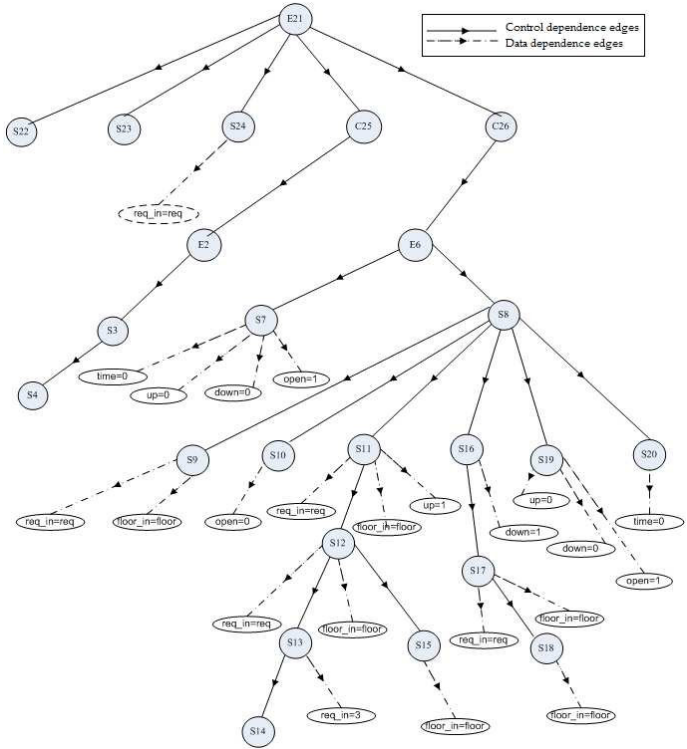

Fig. 4. ClDG of the whole program (Obtained from Figure 2)

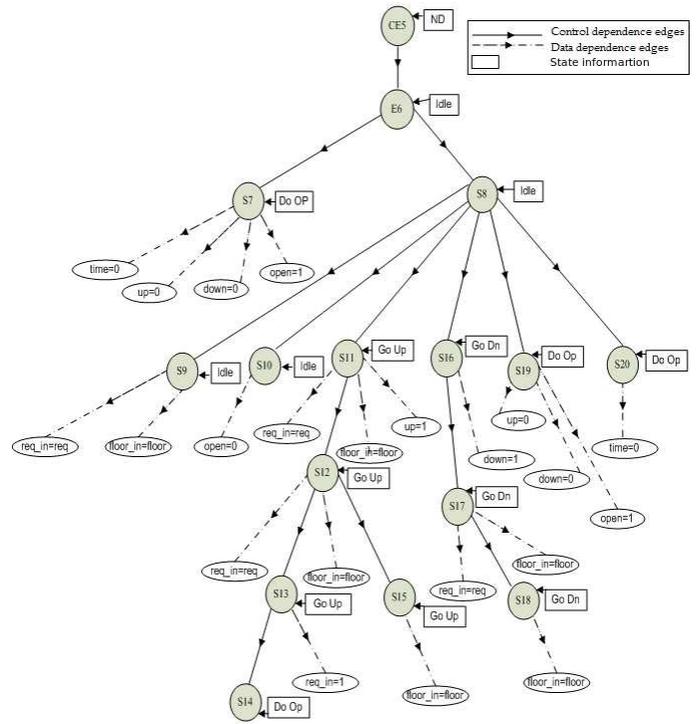

Fig. 5. State information incorporated to ClDG of the class Control

| Initial State | Condition    | Operation (Action) | Final State |
|---------------|--------------|--------------------|-------------|
| Idle          | req == floor | u,d,o,t =0,0,1,0   | Idle        |
| Idle          | req > floor  | u,d,o,t =1,0,0,0   | Going Up    |
| Idle          | req < floor  | u,d,o,t =0,1,0,0   | Going Down  |
| Going Up      | req > floor  | u,d,o,t =1,0,0,0   | Going Up    |
| Going Up      | ! req > floor| u,d,o,t =0,0,1,0   | Door Open   |
| Door Open     | timer < 10   | u,d,o,t =0,0,1,1   | Door Open   |
| Door Open     | ! timer < 10 | u,d,o,t =0,0,1,0   | Idle        |
| Going Down    | req < floor  | u,d,o,t =0,1,0,0   | Going Down  |
| Going Down    | ! req < floor| u,d,o,t =0,0,1,0   | Door Open   |

TABLE I
State transition table generated from Figure 3

### D. ClDG representation of the buggy program

In this section, I will generate the ClDG corresponding to the buggy object oriented program (Figure 2). Along with this representation, I demonstrate another ClDG incorporated with state information of *class Control*, from the program. I choose this class regardless of other two modules of program (i.e. *class RequestResolver* and *void main*).

*1) State information incorporation to ClDG:* From the program shown in Figure 2, it can easily be said that the *class Control* is the main module that controls over the elevator movement. Hence, for simplicity I am interested only to this module of the program. Here I have introduced a critical metric – object state information, into the various nodes of the ClDG of the *class Control*. The similar is shown in the Figure 5. Where each rectangular box is associated with the state information to the corresponding nodes of the ClDG.

### E. Test suite deployment

In this portion, I will feed the buggy program with input test suite. This results in a test suite decision table. This table shows all the combination that can form from a three storied building architecture, capturing its three floors and the corresponding requests. This table shows all the nodes belong to *class Control*. The '+' sign represents the full execution of the respective nodes, while '-' sign tells about the bypassing nodes. Finally the combination of inputs result in the fail or pass mark, shown in Figure 6.

All test cases involving <floor, req> are shown in the Figure 6. From this, I got the clear view of control dependence flow in respect to the nodes of ClDG. The <0, 1> input results in failure, where as others pass. Analyzing the whole test suit decision table, it can promptly be said that input <0, 1> and <1, 2> are closely matched to each other (in respect of control dependence flow).

Fig. 6. Test suite decision table for *class Control*

Fig. 7. State comparison between the failure and passing test cases

Fig. 8. State comparison matrix

## F. State information comparison

Here I present the state information comparison between the above pair of test inputs; the failed input <0, 1> and the passing input <1, 2>. The Figure 7 illustrates the whole process. Nd, Id, Do, Gu, Do represents Not defined, Idle, Door open, Going up, Going down respectively. From the Figure 7, it is clear to understand that the two inputs <0, 1> and <1, 2> differs at node number S13 and S14, in respect of states (*Gu at* S13 *and Do at* S14*)*, shown in rounded form.

## G. State comparison matrix

State comparison matrix is such a matrix that represents the state alignment of failing input to the passing one. The Figure 8 shows the state comparison between <0, 1> and <1, 2>. The left most column presents the state information of passing input (<0, 1>). Whereas the upper most row presents the state information of failing input (<1, 2>). The matrix basically shows a straight line starting from upper left most corner to the lower right most. The line represents the state alignment between these two test inputs. There is a slope in the line, S13 and S14 numbered columns.

## H. Source level bug detection

Now if, I take the information from the above paragraph and search the reason of the misbehavior of the line in the matrix, I can find that a couple of state change occurred during the execution of the code. While investigating this reason, I found that line number 13 of the program involves an *if statement*, which when executes the bug is infiltrated inside the code making it buggy.

```
     int up, down, open, floor=0, req;
CE1  class RequestResolver{
E2       int resolver(){
S3           while (1){
                 ...
                 //Get request from user
S4               cin>>req;
             }
         };
CE5  class Control{
E6       void unitControl(){
S7           int time= up = down = 0, open = 1;
S8           while (1) {
S9               while (req == floor); // Idle
S10              open = 0;
S11              if (req > floor) { up = 1; // Going Up
S12                  while (req != floor) {
S13                      if (req == 1)      Error detected at
S14                          goto P;         line 13 and 14
S15                      floor++; } }
S16              else {down = 1; // Going Down
S17                  while (req != floor)
S18                      floor--; }
S19          P: up = down = 0; open = 1;
                 // Wait for 10000 ms // Door Open
S20              for ( time=0; time < 10000; time++);}
             }
         };
E21  void main()
     {
S22      Control *c = new Control();
S23      RequestResolver *r = new RequestResolver();
S24      while(1) {    // Call concurrently:
C25          r -> resolver();
C26          c -> unitControl();}
     }
```

Fig. 9. Bug localization in source code level

The *if statement* is true when it satisfies the *req==1* condition, resulting the control flow to jump at line number 19. This keeps the door open, making the *Door open* state true. The Figure 9 shows the buggy segment inside the code.

## V. Conclusion

In this paper I have proposed a new methodology for debugging errors in the object oriented programs in source code level. I have given an OSD model as an implementation of object state information into ClDG. I have proven the debugging methodology, incorporating a previously known buggy program into a debugged one. Currently I am busy with the implementation of the OSD model.